# 3D-TDA: Topological feature extraction from 3D images for Alzheimer's disease classification

Faisal Ahmed[†1], Taymaz Akan[†2], Fatih Gelir[2], Owen T. Carmichael[3], Elizabeth A. Disbrow[4], Steven A. Conrad[2] and Mohammad A. N. Bhuiyan[*2]

[1]Department of Mathematical Sciences, University of Texas at Dallas
[2]Department of Medicine, Louisiana State University (LSU) Health Sciences Center, Shreveport
[3]Pennington Biomedical Research Center, LSU
[4]Department of Neurology and Center for Brain Health, LSU Health Sciences Center, Shreveport
faisal.ahmed@utdallas.edu, {taymaz.akan, fatih.gelir}@lsuhs.edu, owen.carmichael@pbrc.edu, {elizabeth.disbrow, steven.conrad, nobel.bhuiyan}@lsuhs.edu

[†]These authors contributed equally to this work.

## Abstract

Now that disease-modifying therapies for Alzheimer's disease (AD) have been approved by regulatory agencies, the early, objective, and accurate clinical diagnosis of AD based on the lowest-cost measurement modalities possible has become an increasingly urgent need. In this study, we propose a novel feature extraction method using persistent homology to analyze structural MRI of the brain. This approach converts topological features into powerful feature vectors through Betti functions. By integrating these feature vectors with a simple machine learning model like XGBoost, we achieve a computationally efficient machine learning (ML) model. Our model outperforms state-of-the-art deep learning models in both binary and three-class classification tasks for ADNI 3D MRI disease diagnosis. Using 10-fold cross-validation, our model achieved an average accuracy of 97.43% and sensitivity of 99.09% for binary classification. For three-class classification, it achieved an average accuracy of 95.47% and sensitivity of 94.98%. Unlike many deep learning models, our approach does not require data augmentation or extensive preprocessing, making it particularly suitable for smaller datasets. Topological features differ significantly from those commonly extracted using convolutional filters and other deep learning machinery. Because it provides an entirely different type of information from ML models, it has the potential to combine topological features with other ML models later on.

## 1 Introduction

Alzheimer's disease (AD) is a brain disorder characterized by the accumulation of abnormal beta amyloid and phosph-tau protein deposits associated with apoptosis and brain atrophy. AD results in decreased memory and other cognitive functions, as well as mood and behavior changes. AD can be characterized by the presence of plaques and tangles, and clinical dementia syndrome. It is preceded in time by a state of having plaques and tangles as well as some cognitive decline, but not frank dementia (called MCI), and preceded further by a state of having plaques and tangles but no identifiable cognitive decline (preclinical AD).. The preclinical stage is characterized by brain, blood, and cerebrospinal fluid abnormalities without outward signs, and AD pathology begins at least 20 years before symptoms appear [1]. The second stage, MCI, involves cognitive impairment confined to a single domain, usually memory. Dementia, the final stage, is characterized by cognitive disturbances in multiple domains, often affecting memory and executive function. Early detection is crucial for successful intervention and management.

AD pathology has been detected using neuroimaging. The importance of structural brain magnetic resonance imaging (MRI) is increasingly highlighted in the early diagnosis of AD. Neuroimaging techniques, such as structural MRI (sMRI), have not only identified specific structural changes in the brains of AD patients but also provided new insights into brain function [2]. sMRI scans capture the three-dimensional shape and structure of brain tissue, enabling detailed analysis and precise volume measurements of various brain components [3].

Early AD is associated with atrophy of the entorhinal cortex, hippocampus, and posterior cingulate gyrus, with a 24% reduction in bilateral hippocampal volume in AD patients relative to normal controls [4]. In addition, frontal and parietal lobes have been shown to exhibit greater atrophy in people with AD compared to mild cognitive impairment. Additionally, [5] introduced a machine-learning framework to differentiate AD patients from those with mild cognitive impairment.

Given the complexity and volume of neuroimaging data, advanced analytical techniques such as machine learning (ML), topological data analysis (TDA), and deep learning have become invaluable in the early detection and diagnosis of Alzheimer's disease. ML algorithms can analyze vast datasets, identifying subtle patterns and features that may be indicative of early-stage AD or MCI. TDA provides a robust framework for extracting features from images [6], capturing the underlying topological structures; however, the application of TDA in imaging, especially in 3D images, is still relatively limited, with fewer studies available. On the other hand, deep learning, with its powerful neural networks, can automatically learn hierarchical features from raw data, improving diagnostic accuracy and aiding in predicting disease progression. Together, these advanced techniques enhance the capability to detect and classify AD at its earliest stages, facilitating timely intervention and better patient outcomes. In this paper, we present a novel approach for diagnosing AD using 3D MRI, leveraging the application of TDA. TDA tools have been successful in medical image analysis by capturing hidden shape patterns within images and generating reliable representations. Topological features extracted provide unique insights into data, enabling discovery of new information and determining relevant features. When combined with suitable machine learning models, these feature vectors enable the development of interpretable, robust, and high-performance machine learning models.

## 2  TDA in Image Processing

TDA represents an innovative method to study complex data by identifying both local and global structures across various scales. It addresses challenges related to data dimensionality, differences in data collection methodologies, and varying scales [7]. Over the past decade, TDA has been widely applied across multiple domains, including image analysis, neurology, cardiology, hepatology, gene-level and single-cell transcriptomics, drug discovery, evolution, and protein structural analysis [8]. By leveraging inherent topological features within images, TDA provides a fresh perspective on image analysis. The ability of TDA to capture hidden patterns in images opens new opportunities for tasks such as image segmentation, object recognition, image registration, and image reconstruction. One commonly utilized tool from TDA in image analysis is persistent homology (PH), which has demonstrated impressive results in pattern recognition for image and shape analysis over the past two decades [9]. In the realm of medical image analysis, PH has proven effective in the analysis of hepatic lesions, histopathology [10], fibrin images, retinal images [11], [12], Chest X-ray images [6], tumor classification, neuronal morphology, brain artery trees [13], fMRI data [14], and genomics data [15]. A thorough review of TDA methods in biomedicine is available in excellent surveys [16].

## 3  Method

In this study, we employ PH as an effective feature extraction method for classification of the stages of AD from MRIs. PH is a key technique in TDA, which allows for systematic exploration of hidden patterns in data by varying a scale parameter. The extracted patterns, known as topological features, and their persistence throughout the filtration process, reveal significant insights into the data's characteristics and organization. We provide a basic overview of PH within the context of image analysis (cubical persistence) for those unfamiliar with the concept. More comprehensive background information and PH applications to other data types, such as point clouds and networks, can be found in[17]. The PH process consists of three main steps: first, the filtration step, where one induces a sequence of cubical complexes from the image data; second, the persistence diagram step, where PH machinery records the evolution of topological features (birth/death times) in the filtration sequence; and finally, the vectorization step, where one can convert these records to a feature vector to be used in suitable ML models.

## 3.1 The Persistent Homology Process

Step 1: Constructing Filtrations

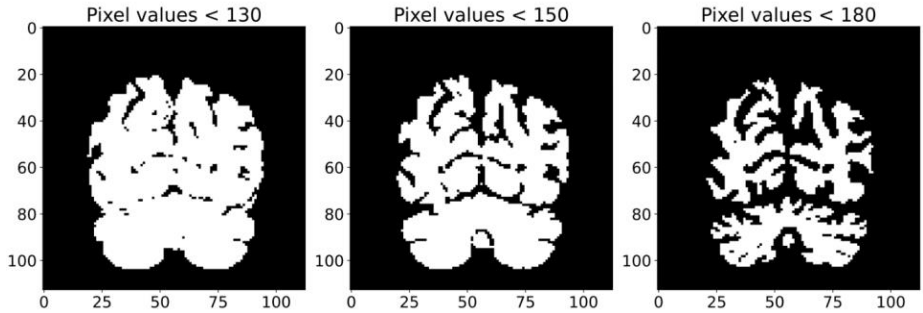

Figure 1: A series of binary images $X_{130}, X_{150}, X_{180}$ are generated from a 2D slice of Alzheimer's disease (AD) using threshold values 130, 150, and 180 respectively.

PH involves tracking the evolution of topological features in a sequence of cubical complexes. For image analysis, this sequence is typically constructed from a nested series of binary images. Given a 3D grayscale image X (e.g., with m × n pixels and a depth of h slices), we use the grayscale values $Y_{ijk}$ of each pixel $\Delta_{ijk} \subset X$. For a sequence of grayscale values $t_1 < t_2 < \cdots < t_N$, we obtain a nested sequence of binary images $X_1 \subset X_2 \subset \cdots \subset X_N$, where $X_n = \{\Delta_{ijk} \subset X \mid Y_{ijk} \leq t_n\}$. This is called sublevel filtration. Alternatively, superlevel filtration can be used by activating pixels in decreasing order of grayscale values. Both methods convey the same information due to Alexander duality.

Step 2: Persistence Diagrams

The second step in PH process involves obtaining persistence diagrams (PDs) for the filtration $X_1 \subset X_2 \subset \cdots \subset X_N$, which represents the sequence of cubical complexes (binary images). PDs are formal summaries of the evolution of topological features in this filtration sequence, represented as collections of 2-tuples $\{(b_\sigma, d_\sigma)\}$, where $b_\sigma$, and $d_\sigma$ denote the birth and death times of the topological features appearing in the filtration. Specifically, if a topological feature σ appears for the first time at $X_{i_0}$, the birth time is $b_\sigma = i_0$ is recorded. If the topological feature σ vanishes at $X_{j_0}$, the death time $d_\sigma = j_0$ is marked. In other words, $PD_k(X) = \{(b_\sigma, d_\sigma) | \sigma \in H_k(\widehat{X}_1)$ for $b_\sigma \leq i \leq d_\sigma\}$, where $H_k(X_i)$ represents the $k^{th}$ homology group of $X_i$, capturing $k -$ dimentional topological features (i.e. $k -$ holes) in the cubical complex $X_i$. For 3D image analysis, meaningful dimensions are $k = 0, k = 1,$ and $k = 2$ corresponding to $PD_0(X), PD_1(X),$ and $PD_2(X)$. For instance, $0 -$ dimensional features are connected components, and $1 -$ dimensional features are loops (holes), $2 -$ dimensional features are cavities (voids). For example, in Figure 2, if a loop σ first appears at the binary image $X_3$ and is filled in at $X_5$, the 2-tuple $(3, 5)$ is added to the persistence diagram $PD_1(X)$. Similarly in Figure 2, if a new connected component appears in $X_1$ and merges with others in $X_4$, $(1,4)$ is added to $PD_0(X)$. For example, in Figure 2, Image X is a 5 × 5 greyscale image. After performing sublevel filtration, we obtain sequence of binary images $X_1 \subset X_2 \subset X_3 \subset X_4 \subset X_5$. Consequently, the diagrams are $PD_0(X) = \{(1, \infty), (1,2), (1,3), (1,3), (1,4), (2,3)\}$ and $PD_1(X) = \{(3,5), (3,5), (4,5)\}$.

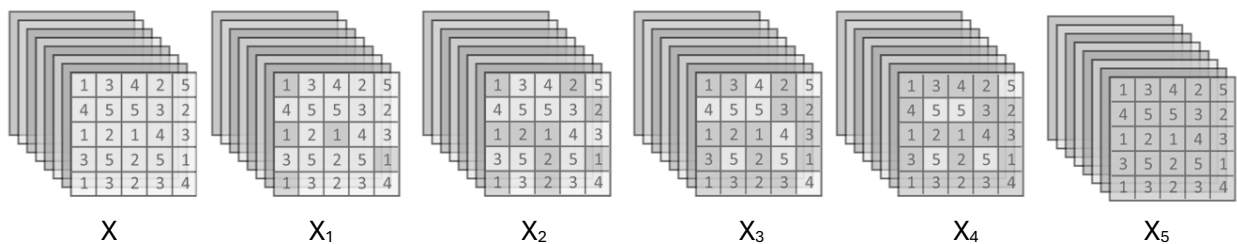

Figure 2: The figure on the left displays an image with dimensions of 5 × 5, showing specific pixel values. The sublevel filtration process involves creating a sequence of binary images, represented as $X_1 \subset X_2 \subset X_3 \subset X_4 \subset X_5$.

Step 3: Vectorization

PDs, being collections of 2-tuples, are not directly usable in machine learning (ML) models. Therefore, vectorization is employed to convert PD information into vectors or functions. A common vectorization method is the Betti function, which tracks the number of 'alive' topological features at each threshold. In particular, the Betti function is a step function with $\beta_0(t_n)$ the count of connected components in the sequential binary image $X_n$, $\beta_1(t_n)$ the number of holes (loops) in $X_n$, and $\beta_2(t_n)$ the number of cavities (voids) in $X_n$. In ML applications, Betti functions are usually taken as a vector of size N with entries $\beta(t_n)$ for $1 \leq n \leq N$, i.e. $\vec{\beta_k}(X) = [\beta_k(t_1) \ldots \beta_k(t_N)]$. The $k^{th}$ Betti curve $\beta_k : [\alpha_1, \alpha_N] \rightarrow \mathbb{Z}$ is an integer valued step function where $\beta_k(\alpha_i)$ is the total # barcodes in $PD_k(X)$ containing $\alpha_i$. For example, from Figure 2, we have $\vec{\beta_0}(X) = [5\ 4\ 2\ 1\ 1]$ and $\vec{\beta_1}(X) = [0\ 0\ 2\ 3\ 0]$.

One can consider TDA as a powerful feature extraction method that captures shape patterns in images, with the Betti vectors $\vec{\beta_0}(X), \vec{\beta_1}(X)$, and $\vec{\beta_2}(X)$ serving as corresponding feature vectors. It's important to note that there are other various methods to convert PDs into vectors in Persistent Homology (PH) (vectorization), such as Persistence Landscapes, Persistence Images, and Silhouettes. Silhouette function $\rho$ is defined as $\rho(X) = \frac{\sum_{j=1}^{N} w_j \Delta_j(t)}{\sum_{j=1}^{N} w_j}$, $t \in [b_j, d_j]$, where $w_j = (d_j - b_j)^p$ is weight. The choice of vectorization method can significantly impact the model's performance depending on the data type.

Generally, topological features with short lifespans are considered topological noise. While other vectorization methods aim to minimize topological noise, Betti functions also consider these short-lived features along with dominant ones. In the case of Alzheimer's 3D MRIs, most topological features have short lifespans, making Betti functions effective in capturing topological patterns. Additionally, among these vectorization methods, Betti functions are the easiest to interpret since they directly count topological features. For these reasons, we utilize Betti functions as the vectorization method in this study.

### 3.2 Topological Features Extraction from Alzheimer's 3D MRIs

The flowchart (Figure 3) is a summary of our model. Since all Alzheimer's 3D MRIs consist of sequential 2D slices and are grayscale images, they dictate a gray scale filtration method for our persistent homology (PH) approach. Using the greyscale pixel values for any Alzheimer's 3D MRI X, we define a sublevel filtration as described in the previous section. While grayscale values range from 0 (black) to 255 (white), we chose the number of thresholds as $N = 100$ in our filtration step, as further increasing the number of thresholds did not improve the performance of our model. In other words, we normalized the grayscale interval $[0, 255]$ to $[0, 100]$. After defining the filtration $X_1 \subseteq X_2 \subseteq \ldots \subseteq X_{100}$, we obtain the persistent diagrams $PD_k(X)$ of each 3D MRI X for dimensions $k = 0, 1, 2$. Since MRI are 3D, only $k = 0, 1, 2$ are meaningful dimensions for PH.

After obtaining persistent diagrams, we convert them into feature vectors as explained in the previous section. In this vectorization step, several choices can be used, such as Betti functions, Silhouettes, or Persistence Images. Given that most topological features have short lifespans, Betti functions were the natural choice as they provide the count of topological features at a given threshold. Furthermore, for interpretability, we use Betti functions in our models. Thus, we convert $PD_0(X), PD_1(X)$, and $PD_2(X)$ into the corresponding Betti functions $\beta_0(X), \beta_1(X)$, and $\beta_2(X)$ as our feature vectors (Figure 3 - Step 2,3). Therefore, our topological feature extraction process produces 300 features for any 3D MRI X. To use machine learning tools more effectively, we generate functions (topological summaries) from these persistence diagrams.

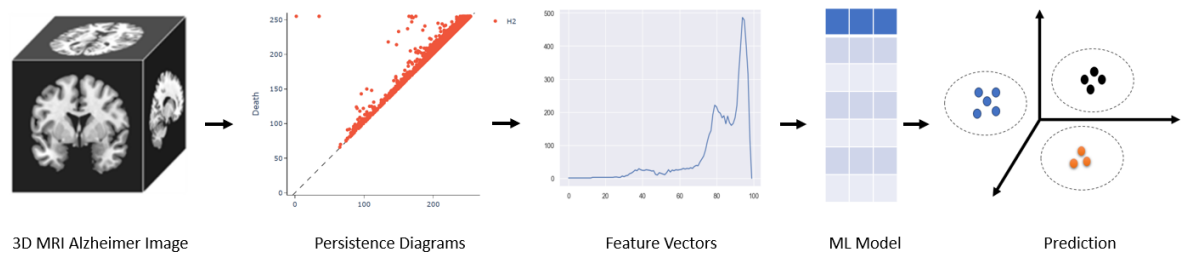

Figure 3: Flowchart of our model: For any 3D MRI of AD, we first generate persistence diagrams using the grayscale values ranging from 0 to 255. From these persistence diagrams, we extract topological feature vectors by using Betti functions. Instead of feeding the original images to our machine learning models, we use these feature vectors as input. This approach yields highly accurate classification results.

### 3.3 ML Model

After obtaining our topological feature vectors, the final step is to apply ML tools to these topological fingerprints. To keep our model computationally feasible, we applied XGBoost to classify these Betti vectors. No data augmentation or preprocessing was used for our model. The model uses topological feature vectors, and our feature extraction method is invariant under rotation, flipping and other common data augmentation techniques. This makes our model computationally very feasible and can easily be applied to very large datasets.

## 4 Dataset

The Alzheimer's disease Neuroimaging Initiative (ADNI) database (adni.loni.usc.edu) was the source of the data utilized in this study [18], [19]. In 2003, the ADNI was established as a public-private partnership under the leadership of Principal Investigator Michael W. Weiner, MD. The primary objective of ADNI has been to determine whether the progression of mild cognitive impairment (MCI) and early Alzheimer's disease (AD) can be monitored through the combination of serial magnetic resonance imaging (MRI), positron emission tomography (PET), clinical and neuropsychological assessment, and other biological markers. Using T1-weighted 3D MRI scans from ADNI, we conducted binary and multiclass classification. The models were trained and tested on subjects who underwent scans at screening, as well as at 6 months, 1 year, and 18 months (MCI only), and 2 and 3 years (normal and MCI only) (ADNI1: Complete 3Yr 3T data). We performed 10-fold cross-validation on the dataset. In order to evaluate the proposed method, we conducted extensive experiments on binary (NC/AD) and multi-classification tasks (NC/MCI/AD). Moreover, only the middle 50 slices were considered the model input.

## 5 Experimental Results and Setup

The Alzheimer's Disease Neuroimaging Initiative (ADNI) dataset is widely used for evaluating machine learning models in the field of neuroimaging, specifically for classifying AD patients. Several models have been applied to this dataset with varying image classes and performance metrics. In this study, we used XGBoost to classify the Betti vectors extracted from 3D MRI provided by ADNI dataset [18], [19]. In order to evaluate the proposed method, we conducted extensive experiments on binary (NC/AD) and multi-classification tasks (NC/MCI/AD). AD usually has three stages. Preclinical (NC) is the first stage, with brain and blood without symptoms. AD pathology may start 20 years before symptoms. The second stage of the disease is mild cognitive impairment (MCI), which affects one cognitive domain, usually memory. The final stage of the disease, dementia (AD), is a cognitive disturbance in multiple domains, often memory and executive function, that interferes with daily life [20], [21]. For binary classification, we merged the MCI and AD disease classes into a single class. The model was constructed and evaluated using a 10-fold cross-validation strategy to ensure robustness and generalizability of the results. To enhance our model's performance in terms of accuracy and computational efficiency, we did parametric tuning and feature selection methods. For feature selection, we used the

**SelectFromModel** function from scikit-learn. We first assigned importance scores to each feature and then sorted them in descending order based on these scores. Features with importance values below the specified threshold were considered unimportant and subsequently ignored. We got the best model performance using XGBoost with the following parameters: colsample_bytree = 0.3, learning_rate = 0.2, max_depth = 7, and n_estimators = 500. These parameters resulted in the highest average accuracy. To comprehensively evaluate the model's performance, we computed precision, recall, accuracy, AUC (Area Under the Curve), and F1-score for each fold, and subsequently calculated their mean values. For multi-class classification, we computed the Area Under the Receiver Operating Characteristic Curve (ROC AUC) using the one-vs-rest (OvR) approach.

## 6 Results

In this section, we present the performance of our model on Alzheimer's 3D. It is important to note that $\beta_0$ refers to 0-dimensional topological features (components), $\beta_1$ refers to 1-dimensional topological features (loops or holes), and $\beta_2$ refers to 2-dimensional topological features (cavities or voids) extracted from Alzheimer's 3D MRIs. $\beta_1 + \beta_2$ combines $\beta_1$ (100-dimensional) and $\beta_2$ (100-dimensional) features, totaling 200-dimensional features. $\beta_0 + \beta_1 + \beta_2$ combines all three feature sets, totaling 300-dimensional features. We will now compare the performance of our model with recent deep learning models for both binary and 3-class classification settings.

### 6.1 Binary Classification:

The performance of various models on the ADNI dataset for 2-class classification highlights differences in their ability to capture relevant features in medical imaging. As shown in Table 1, the High-Frequency Component Network (HFCN) achieves an accuracy of 90.30%, sensitivity of 82.40%, and specificity of 96.5% using a training dataset on ANDI-1 and test dataset on ADNI-2. The 3D DenseNet, employing a five-fold cross-validation on 119/233/97 image classes, reaches an accuracy of 88.90%, with a sensitivity of 86.60% and specificity of 90.80%. A 3D ResNet model, using a 90:10 train-test split with 457/808/346 image classes, achieves an impressive accuracy of 94.00%, although sensitivity and specificity metrics are not reported. The 3D-CNN model, with a 67:33 split on 330/299/299 image classes, performs at 93.20% accuracy, 95.00% sensitivity, and 89.80% specificity. A 2.5D-CNN applied with a 10-fold cross-validation on 209/401/188 image classes achieves 79.90% accuracy, 84.00% sensitivity, and 74.80% specificity. The proposed model, using a 10-fold cross-validation on 129/145/77 image classes, stands out with a remarkable accuracy of 97.43%, sensitivity of 99.09%, and specificity of 95.43%, highlighting its potential effectiveness for Alzheimer's classification tasks.

Figure 4a shows the average confusion matrix (percentage) for the proposed model, highlighting its performance in terms of true positives, true negatives, false positives, and false negatives. This visualization confirms the robustness and accuracy of the proposed model in binary classification tasks for Alzheimer's 3D MRIs.

Table 1. Results for binary disease patient classification on ADNI

| Method | Image Classes | Train/Test | Class | Acc | Sen | Spe |
| --- | --- | --- | --- | --- | --- | --- |
| HFCN [22] | 429/–/858 | - | 2 | 90.30 | 82.40 | 96.5 |
| 3D DenseNet [23] | 119/233/97 | 5-Fold | 2 | 88.90 | 86.60 | 90.80 |
| 3D ResNet [24][19] | 457/808/346 | 90/10 | 2 | 94.00 | - | - |
| 3D-CNN [25] | 330/299/299 | 67/33 | 2 | 93.20 | 95.00 | 89.80 |
| 2.5D-CNN [26] | 209/401/188 | 10-Fold | 2 | 79.90 | 84.00 | 74.80 |
| Proposed Model | 129/145/77 | 10-Fold | 2 | **97.43** | **99.09** | 95.43 |

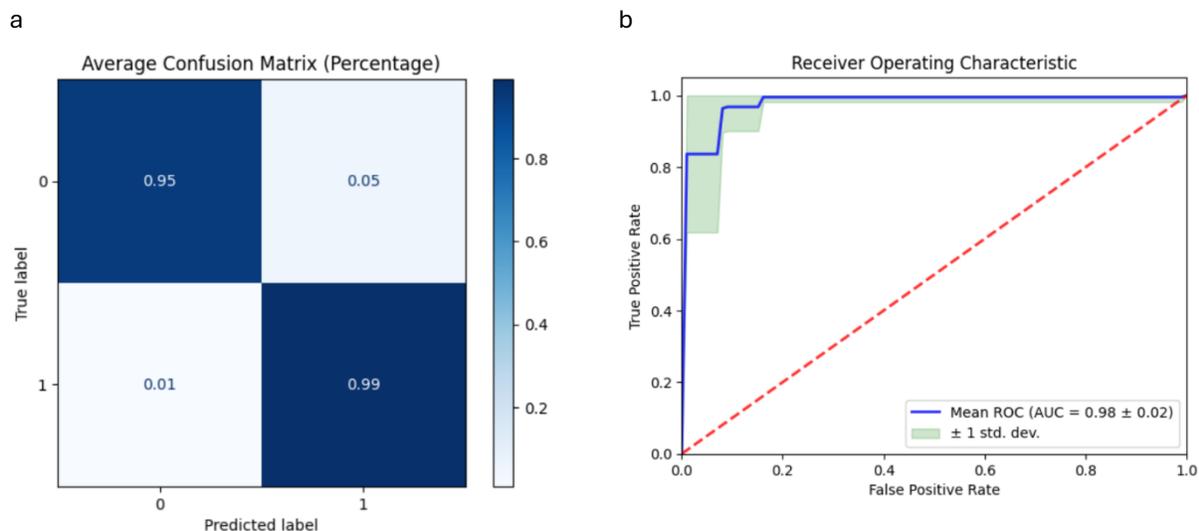

**Figure 4 a** Average Confusion matrices for multiclass disease patient classification on ADNI. The confusion matrix shows that the model correctly identifies 95% of true negatives and 99% of true positives with 5% and 1% false positives and negatives, respectively. **b** Average Receiver Operating Characteristic (ROC) Curve for binary Classification using 10-fold cross-validation.

Figure 4b illustrates the Receiver Operating Characteristic (ROC) curve for the proposed model, showcasing its performance in distinguishing between the positive and negative classes. The ROC curve plots the true positive rate (sensitivity) against the false positive rate (1-specificity) at various threshold settings.

### 6.2 Multi-Class Classification:

In the realm of 3-class classification using the ADNI dataset, several advanced models have demonstrated their effectiveness. Table 2 reveals that the DemNet approach, applied with a 70:30 train-test split across 300/300/300 image classes, achieves a high overall accuracy of 91.85%, though specific precision, recall, and F1-score metrics are not reported. The Natural Image base model, trained on 755/2282/1278 images with a 90:10 split, attains an accuracy of 85.00% with recall of 90.51%. Similarly, the MRI-base model shows an accuracy of 78.20% and recall of 78.07%. The 3D ResNet model, employing a 90:10 split on 457/808/346 image classes, achieves an accuracy of 87%, though other performance metrics are not detailed. The 3D VGGNet model, with a 90:10 split on 207/215/193 image classes, reaches an accuracy of 91.13%. Notably, the proposed model, evaluated using 10-fold cross-validation on 129/145/77 image classes, excels with a precision of 95.97%, recall of 94.98%, accuracy of 95.47%, and an impressive F1-score of 95.24%, showcasing its superior performance in the classification of AD across three classes.

Table 2. Results for multiclass (Three classes) disease patient classification on ADNI

| Method | Image Classes | Train/Test | Classes | Prec | Rec | Acc | F1-score |
|---|---|---|---|---|---|---|---|
| DemNet [27] | 300/300/300 | 70/30 | 3 | - | - | 91.85 | - |
| Nat.Img base [28] | 755/2282/1278 | 90/10 | 3 | - | 90.51 | 85.00 | - |
| 2D CNN [28] | 755/2282/1278 | 90/10 | 3 | - | 78.07 | 78.20 | - |
| 3D ResNet [24] | 457/808/346 | 90/10 | 3 | - | - | 87 | - |
| 3D VGGNet [29] | 207/215/193 | 90/10 | 3 | - | - | 91.13 | - |
| Proposed Model | 129/145/77 | 10-Fold | 3 | **95.97** | **94.98** | **95.47** | **95.24** |

Figure 5a presents the average confusion matrix (percentage) for the multi-class classification task, visualizing the performance of the proposed model across three classes. The matrix provides insights into the true positive, true negative, false positive, and false negative rates for each class. The values along the diagonal represent the percentage of correctly predicted labels for each class, while the off-diagonal values indicate the misclassification rates. For class 0 (NC), the model correctly identifies 95% of instances, with 5% misclassified as class 1 and 1% as class 2. For class 1 (MCI), the model achieves a 98% correct classification rate, with 1% misclassified as class 0 and 1% as class 2. For class 2 (AD), the model correctly identifies 92% of instances, with 4% misclassified as class 0 and 4% as class 1. This confusion matrix illustrates the high performance and reliability of the proposed model in distinguishing between the three classes, with a strong emphasis on correctly predicting true labels while maintaining low misclassification rates

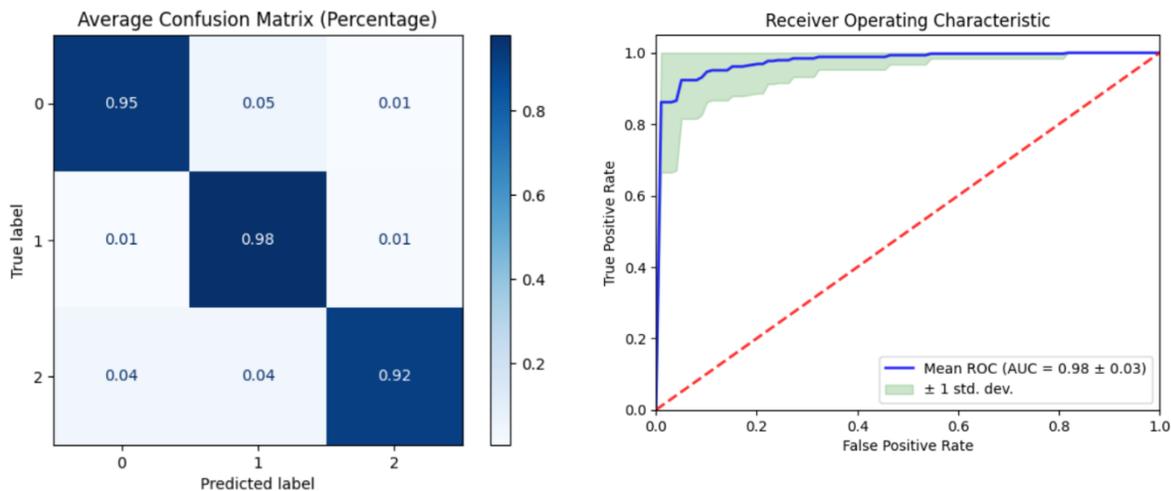

**Figure 5. a** Average Confusion matrices for multiclass disease patient classification on ADNI. The model correctly identifies 95% of class 0 (NC) instances, misclassifying 5% as class 1 and 1% as class 2. Class 1 (MCI) is classified 98% correctly, with 1% misclassified as class 0 and 1% as class 2. For class 2 (AD), the model correctly identifies 92% of instances, misclassifying 4% as class 0 and 4% as class 1. **b** Average Receiver Operating Characteristic (ROC) Curve for Multi-Class Classification using 10 fold cross-validation.

Figure 5b illustrates the Receiver Operating Characteristic (ROC) curve for the proposed model in a multi-class classification task. The ROC curve plots the true positive rate (sensitivity) against the false positive rate (1-specificity) at various threshold settings.

## 7 Discussion

The application of persistent homology (PH) as a feature extraction method in 3D MRI imaging for Alzheimer's disease (AD) presents a novel approach that leverages the strengths of topological data analysis (TDA). Our study demonstrates that PH can effectively capture topological features, such as loops and cavities, within 3D MRI data, providing insightful patterns that contribute to accurate classification of AD stages. The key advantage of using PH in this context is its ability to maintain interslice dependencies, which is crucial for preserving the spatial and structural integrity of 3D images. As our figures suggest, our topological descriptors are highly effective in distinguishing between different stages of AD. Remarkably, even without employing deep learning techniques, our model surpasses state-of-the-art models in benchmark datasets for AD classification (Table 1 and 2). Our results clearly demonstrate the remarkable capability of TDA in generating powerful feature vectors from 3D MRI images. However, the number of studies utilizing TDA for 3D image analysis, particularly in medical imaging, remains limited. While our method shows significant promise, it is important to acknowledge that TDA's application in this area is still in its early stages. Despite this, the integration of these topological descriptors with machine learning models has proven highly effective in our experiments. The ability to transform complex 3D MRI data into manageable and interpretable topological feature vectors addresses the challenges of computational feasibility and interpretability faced by traditional ML and DL models. This transformation not only makes the data more manageable but also retains essential structural information that is often lost in other preprocessing methods. Our experiments revealed that the proposed method, integrating PH-derived features with XGBoost, significantly outperformed state-of-the-art DL models in both binary and multi-class classification tasks on the ADNI dataset. The use of Betti functions as feature vectors proved particularly effective, providing a robust and interpretable representation of the topological features present in the MRI images. Nevertheless, our results also indicate that there is room for improvement, particularly when it comes to integrating these topological features with advanced deep learning models. By seamlessly combining these descriptors with deep learning, we can potentially enhance the effectiveness of clinical decision support systems for AD diagnosis.

## 8 Conclusion

Our study presents a pioneering approach to Alzheimer's disease diagnosis using 3D MRI images by leveraging the strengths of persistent homology and topological data analysis. The integration of PH with machine learning, specifically XGBoost, has demonstrated superior performance in classifying AD stages, highlighting the potential of this method to enhance early detection and diagnosis. The ability of PH to maintain interslice dependencies and provide a comprehensive analysis of 3D structures is a significant advancement in neuroimaging. The results underscore the importance of further exploring and developing TDA methods for 3D medical imaging, as they offer a promising avenue for improving diagnostic accuracy and patient outcomes. Future work should focus on expanding the application of TDA in medical imaging, exploring other vectorization methods, and integrating these approaches with more advanced ML and DL models. Additionally, large-scale studies and clinical trials will be necessary to validate the effectiveness and reliability of these methods in real-world settings. By continuing to refine and enhance these techniques, we can contribute to the development of more effective clinical decision support systems for Alzheimer's disease and other neurological disorders.

### Acknowledgement

This work was supported by Institutional Development Award (IDeA) from the National Institutes of General Medical Sciences of the NIH under grant number P20GM121307 to MANB, and the project was partially supported by the Ike Muslow, MD Endowed Chair in Healthcare Informatics of LSU Health Sciences Center Shreveport.